\documentstyle[12pt]{article}

\def\bea{\begin{eqnarray}}
\def\eea{\end{eqnarray}}

\def\nn{\nonumber}

\def\xibar{\bar\xi}

\def\K1{{\cal K}_{\bf 1}}
\def\Q20{{\cal Q}_{\bf 20'}}
\def\cK{{\cal K}_{\bf 1}}
\def\cN{{\cal N}}
\def\calQ{{\cal Q}}
\def\calK{{\cal K}}

\newbox\SlashedBox
\def\fs#1{\setbox\SlashedBox=\hbox{#1}
\hbox to 0pt{\hbox to 1\wd\SlashedBox{\hfil/\hfil}\hss}{#1}}
\def\hboxtosizeof#1#2{\setbox\SlashedBox=\hbox{#1}
\hbox to 1\wd\SlashedBox{#2}}

\def\ms#1{\setbox\SlashedBox=\hbox{$#1$}
\hbox to 0pt{\hbox to 1\wd\SlashedBox{\hfil/\hfil}\hss}#1}

\newcommand{\tr}{{\rm tr}}
\newcommand{\ie}{{\em i.e.}}
\newcommand{\eg}{{\em e.g.}}

\newcommand{\be}{\begin{equation}}
\newcommand{\ee}{\end{equation}}
\newcommand{\ba}{\begin{eqnarray}}
\newcommand{\ea}{\end{eqnarray}}

\begin{document}
\date{today}
\thispagestyle{empty}

\begin{flushright}

CERN-TH/2001-096\\
DAMTP-2001-31 \\
ROM2F/01/10
\end{flushright}

\vspace{.5cm}

\begin{center}
{\LARGE {\bf Properties of the Konishi multiplet\\
in ${\cal N}=4$ SYM theory  \rule{0pt}{25pt} }} \\
\vspace{.8cm} \
{Massimo Bianchi$^{\:a \,*}$, Stefano Kovacs$^{\:a}$,
Giancarlo Rossi$^{\:b\,*}$ and Yassen S. Stanev$^{\:c\,\dagger}$} \\
\vspace{0.5cm}

{$^a$ {\it D.A.M.T.P., University of Cambridge}} \\
{{\it Wilberforce Road, Cambridge, CB3 0WA, UK}}\\
\vspace*{0.3cm}
{$^b$ {\it TH Division, CERN}} \\
{{\it 1211 Gen\`eve 23, CH}}\\
\vspace*{0.3cm}
{$^c$ {\it Dipartimento di Fisica, \ Universit{\`a} di Roma \
``Tor Vergata''}} \\  {{\it I.N.F.N.\ -\ Sezione di Roma \
``Tor Vergata''}} \\ {{\it Via della Ricerca  Scientifica,}}
{{\it 00133 \ Roma, \ ITALY}} \\

\end{center}

\vspace{.5cm}

\begin{abstract}
We study perturbative and non-perturbative properties of the
Konishi multiplet in ${\cal N}=4$ SYM theory in $D=4$ dimensions. We
compute two-, three- and four-point Green functions with single and
multiple insertions of the lowest component of the multiplet, 
${\cal{K}}_{\bf 1}$, and of the lowest component of the supercurrent
multiplet, ${\cal Q}_{\bf 20^\prime}$. These computations require a
proper definition of the renormalized operator, ${\cal K}_{\bf 1}$, 
and lead to an independent derivation of its anomalous dimension. 
The  O($g^2$) value found in this way is in agreement with previous
results. We also find that instanton contributions to the above
correlators vanish.

{}From our results we are able to identify some of the 
lowest dimensional gauge-invariant composite operators contributing to 
the OPE of the correlation functions we have computed. We thus confirm
the existence of an operator belonging to the representation ${\bf
20^\prime}$, which has vanishing anomalous dimension at order $g^2$
and $g^4$ in perturbation theory as well as at the non-perturbative
level, despite the fact that it does not obey any of the known
shortening conditions.
\end{abstract}

\vspace{1.0cm}
\noindent
\rule{6.5cm}{0.4pt}

\noindent {\footnotesize ${}^{*}$~On leave of absence from Dipartimento
di Fisica, Universit{\`a} di Roma ``Tor Vergata'', Via della Ricerca
Scientifica,
00133 Roma, ITALY.}

\noindent {\footnotesize ${}^{\dagger}$~On leave of absence from
Institute for
Nuclear Research and Nuclear Energy, Bulgarian Academy of Sciences,
BG-1784, Sofia, Bulgaria.}
\newpage

\setcounter{page}{1}

\section{Introduction}

${\cN} = 4$ supersymmetric Yang-Mills theory in $D=4$ is a very interesting
and pedagogically useful theory. It is completely determined by the
choice of the gauge group and is known to be ``finite''~\cite{finite}. All the
couplings of the theory are related to the gauge coupling, which has
vanishing $\beta$ function both perturbatively and non-perturbatively.
In the superconformal phase (at vanishing scalar {\it vev}'s), the spectrum of
gauge invariant composite operators is very rich. They build up 
representations of the supergroup $SU(2,2|4)$ that contains, as purely 
bosonic symmetries, the conformal group, $SO(4,2)$, and the R-symmetry 
group, $SO(6)\approx SU(4)$.

The lowest (non-trivial) of such representations, the singleton, is made up
by the 8+8 bosonic and fermionic components which are in correspondence with 
the fundamental fields of the theory. It has gauge invariant components only 
in the abelian case. In the non-abelian case, gauge invariant operators must 
be at least bilinears in the fundamental fields. The simplest gauge invariant
supermultiplet is that of the ${\cN}=4$ supercurrent, which comprises 
128 bosonic and as many fermionic operators. The ``lowest lying'' components 
of the multiplet are the so-called chiral primary operators (CPO's), $\Q20$, 
belonging to the representation ${\bf 20'}$ of $SU(4)$~\footnote{$SU(4)$ 
representations are classified by a triplet of Dynkin labels 
$[\ell,m,n]$~\cite{sla}. In particular the Dynkin labels of the 
representation ${\bf 20'}$ are $[0,2,0]$.}. The operators $\Q20$, as well 
as their superconformal partners (among which the stress-tensor, 4 
supercurrents, 15 axial currents, $\cdots$), are protected against quantum 
corrections of their (conformal) dimensions, since the supercurrent 
multiplet is (ultra) short.

All multiplets built starting from CPO's with Dynkin labels of the form
$[0,\ell,0]$ and conformal dimension $\Delta_0=\ell\ge 2$ are short
multiplets. Relying on previous studies of unitary irreps of
$SU(2,2|{\cal{N}})$~\cite{dobpet}, further multiplet shortening conditions
have been identified in ref.~\cite{AFZ} and classified in~\cite{FS}. In
particular the shortening of the multiplet built starting from the
scalar operator of dimension $\Delta_0 = 4$ in the representation ${\bf 84}$ of
$SU(4)$, which has Dynkin labels $[2,0,2]$~\footnote{Actually multiplets
whose lowest scalar components have conformal dimension $\Delta_0=\ell+2k$ 
and Dynkin labels $[k,\ell,k]$ are short~\cite{AFZ,FS}.}, nicely
fits with the results obtained in~\cite{bkrs1,bkrs2}, where the precise field
theoretical definition of this operator has been given. The problem with
this and similar cases is that quantum corrections induce mixing among (bare)
operators with the same quantum numbers and a basis of (independent)
operators with well-defined conformal dimensions has to be identified. In the 
case at hand the relevant mixing involves operators of the so-called ${\cN}=4$
Konishi multiplet. The properties of its ${\cN} =1$ submultiplet (to which
we will refer in the following as the ${\cN} =1$ Konishi multiplet) have
been studied extensively after the discovery of the Konishi
anomaly~\cite{kenkon} (see \eg~\cite{akmrv} for a ``pre--AdS/CFT''
review). There has been a renewed interested in the subject in the light
of the proposed AdS/CFT correspondence~\cite{jm,gkp,w,magoo}, because, as
first observed in~\cite{AF}, the Konishi multiplet is a long multiplet that
corresponds to the first string level in the spectrum of type IIB
excitations around the ${\rm AdS}_5\times S^5$ background.

The purpose of this paper is to study the peculiar perturbative and
non-perturbative properties of the ${\cN} =4$ Konishi multiplet.
We will start by constructing the multiplet, taking in due account
the terms induced by the Konishi anomaly. Following~\cite{AF},
we decompose the multiplet in terms of ${\cN} =1$ submultiplets. We will
then compute correlation functions involving the lowest dimensional
scalar operator of the multiplet, $\cK$, which is an $SU(4)$
singlet with (naive) conformal dimension $\Delta_0=2$. Since it has a
non-vanishing anomalous dimension, these computations require a careful
definition of $\cK$, as a finite, gauge invariant, renormalized
operator.

We will compute to O($g^2$) two-, three- and four-point correlators
involving $\cK$ and/or the lowest scalar CPO's of the supercurrent 
multiplet. If complemented with what is known to O($g^4$) from 
ref.~\cite{bkrs2,aeps}, these calculations, besides confirming the results 
known both at O($g^2$)~\cite{ans,bkrs1} and at O($g^4$)~\cite{bkrs2} on 
the $\cK$ anomalous dimension, allow us to identify some of the operators 
contributing to the OPE of these Green functions and compute their anomalous 
dimensions. This is done by exploiting certain rigorous bounds, that 
can be derived on general grounds on the O($g^2$) corrections to 
the anomalous dimensions of operators contributing to the OPE of 
four-point Green functions. We are able to show in this way at 
O($g^4$), as in~\cite{aeps},
the vanishing of the anomalous dimension of an {\it a priori} 
unprotected scalar operator of dimension $\Delta_0=4$, belonging to the 
representation ${\bf 20'}$, which was already shown to be 
zero at O($g^2$) and non-pertubatively~\cite{afp,afp2}.

We will then consider non-perturbative instantonic contributions to the
same correlation functions and to our surprise we find vanishing results.
This leads us to conclude that not only $\cK$ receives no instantonic
contributions to its anomalous dimension, but it tends to display a much
larger ``inertia'' to non-perturbative corrections. Our results imply
the vanishing of non-perturbative corrections to all presently studied
trilinear couplings involving components of the Konishi multiplet.

The plan of the paper is as follows. After fixing our notations in
Sect.~2  and giving the relevant components of the Konishi
supermultiplet in Sect.~3, in Sect.~4 we construct the properly
renormalized expression of ${\cal K}_{\bf 1}$ and compute to O($g^2$)
its two-, three- and four-point Green functions. We also report
explicit formulae for the four-point Green functions with single and
multiple insertions of $\calK_{\bf 1}$ and the lowest component
operators belonging to the supercurrent multiplet, $\Q20$. In Sect.~5, 
we then derive a set of inequalities that anomalous 
dimensions of operators exchanged in intermediate channels must
satisfy for consistency.
With these results and the knowledge we have from the O($g^4$)
calculation of the Green functions of four $\Q20$ operators, we show 
in Sect.~6 what kind of information is possible to extract 
about the anomalous dimensions of the composite operators of naive 
dimension $\Delta_0=4$, belonging to the representation ${\bf 20'}$
and to the singlet. An interesting corollary of this analysis is that
we are able to extend to O($g^4$), as in~\cite{aeps}, 
the observation, made in ref.~\cite{afp} 
to O($g^2$), that there exists an operator in the representation ${\bf 20'}$ 
which has vanishing anomalous dimension, despite the
fact that it does not obey any of the known shortening conditions. 
At the same time we confirm the known results on the anomalous 
dimension of the Konishi multiplet.
The instanton  contributions to the Green functions considered in the
previous sections are  computed in Sect.~7 and shown to vanish.
Conclusions and an outlook of future lines of investigation can be
found in Sect.~8.

\section{Notations and conventions}

The field content of ${\cal N}=4$ SYM~\cite{n4sym} comprises a
vector, $A_{\mu}$, four Weyl spinors, $\psi^{A}$ ($A=1,2,3,4$), and
six real scalars, $\varphi^{i}$ ($i=1,2,\ldots,6$), all in the adjoint
representation of the gauge group, $SU(N)$. In the ${\cal N}=1$ approach
that we shall follow the fundamental fields can be arranged into a
vector superfield, $V$, and three chiral superfields, $\Phi^{I}$ ($I=1,2,3$).
The six real scalars, $\varphi^{i}$, are combined into three complex
fields, $\phi^{I}$ and $ \phi^{\dagger}_{I}$ that are the lowest
components of the chiral and antichiral superfields, $\Phi^{I}$ and
$\Phi^{\dagger}_{I}$, respectively. Three of the Weyl fermions, $\psi^{I}$, 
are the spinors of the chiral multiplets. The fourth spinor, 
$\lambda = \psi^{4}$, together with the
vector, $A_{\mu}$, form the vector multiplet. In this way only an
$SU(3)\otimes U(1)$ subgroup of the full $SU(4)$ R-symmetry is manifest.

The complete ${\cal N}=4$ SYM action in the ${\cal N}=1$ superfield
formulation has a non-polynomial form. A gauge fixing term
must be added to the classical action. We shall use the Fermi-Feynman
gauge, as it makes corrections to the propagators of the fundamental
superfields vanish at order $g^{2}$~\cite{finite,zanon,kov}. Actually a 
stronger result has been proved in these papers, namely the vanishing of 
the anomalous dimensions of the fundamental fields up to O($g^4$). With 
the Fermi-Feynman gauge choice the terms relevant for the calculation of 
the Green functions we are interested in are
\begin{eqnarray}
S &=& \int d^{4}x\:d^{2}\theta d^{2}\bar\theta\,
\left\{\rule{0pt}{18pt}
V^{a} \Box  V_{a}  - \Phi^{a\dagger}_{I}\Phi^{I}_{a} - 2 ig
f_{abc}{\Phi^{\dagger}}^{a}_{I} V^{b}\Phi^{Ic} + 2 g^{2} f_{abe}
f_{ecd}{\Phi^{\dagger}}^{a}_{I}
V^{b} V^{c} \Phi^{Id}  \right.  \nonumber \\
&& \left. - \frac{ig \sqrt{2}}{3!} f^{abc} \left[
\varepsilon_{IJK}
\Phi_{a}^{I} \Phi_{b}^{J} \Phi_{c}^{K} \delta^{(2)}({\overline \theta})
- \varepsilon^{IJK} \Phi^{\dagger}_{aI}\Phi^{\dagger}_{bJ}
\Phi^{\dagger}_{cK}\delta^{(2)}(\theta) \right] +
\ldots \right\} \, ,
\label{actionsuper}
\end{eqnarray}
where $f_{abc}$ are the structure constants of the gauge group. As neither
the cubic and quartic vector interactions nor the ghost terms will contribute
to the calculations we will present in this paper, we have omitted them
in eq.~(\ref{actionsuper})~\footnote{Unlike what we have done in
refs.~\cite{bkrs1} and~\cite{bkrs2}, in this paper we use a more
standard definition of $g$. To compare formulae of~\cite{bkrs1}
and~\cite{bkrs2} with the present ones, one has to replace $g$
there, with $2g$. We thank B. Eden and H. Osborn for pointing out
to us that our notation was at odds with the standard one.}.

Since all superfields are massless, their propagators have an equally
simple form in momentum and in coordinate space and thus we choose to
work in the latter, which is more suitable for the study of
conformal field theories. In Euclidean coordinate space one finds
\be
\langle \Phi^{\dagger}_{Ia} (x_i,\theta_i, \bar \theta_i)
\Phi^J_b (x_j,\theta_j, \bar \theta_j) \rangle =
 {{\delta_{I}}^{J} \delta_{ab} \over 4 \pi^2}
{\rm e}^{\left( \xi_{ii} +\xi_{jj} -2 \xi_{ji} \right) \cdot \partial_j}
{1\over x_{ij}^2} \, ,
\label{propfi}
\ee
\be
\langle V_a (x_i,\theta_i, \bar \theta_i)
V_b (x_j,\theta_j, \bar \theta_j) \rangle =
- {\delta_{ab} \over 8 \pi^2}
{\delta^{(2)}(\theta_{ij}) \delta^{(2)}(\bar \theta_{ij}) \over x_{ij}^2} \, ,
\label{propv}
\ee
where
$x_{ij} = x_{i}-x_{j}$, $\theta_{ij}= \theta_{i}- \theta_{j}$,
 $\xi^{\mu}_{ij}= \theta_i^{\alpha} \sigma^{\mu}_{\alpha {\dot
\alpha}} \bar \theta_j^{\dot \alpha}$.

\section{The ${\cN} = 4$ Konishi multiplet}

The ${\cN} = 4$ Konishi multiplet is a long multiplet of the superconformal
group, $SU(2,2|4)$. Its lowest component, $\cK$, is a scalar operator of
(naive) conformal dimension $\Delta_0=2$, which is a singlet of the $SU(4)$ 
R-symmetry group. The highest spin component of the Konishi multiplet is
a classically conserved spin 4 current, \ie~a 4 index totally symmetric tensor,
of naive dimension $\Delta_{0} = 6$, 
which is also a singlet of $SU(4)$. 

The (naive) definition of $\cK$ is

\be
\cK(x)\Big{|}_{\rm naive}  = {1\over 2}
\sum_{i=1}^{6} : \tr(\varphi^{i}(x)\varphi^{i}(x)):\, ,
\label{K1naivest}
\ee
where the trace is over colour and the symbol $:$ stands for normal ordering.
As usual, normal ordering means subtracting the operator {\it vev} or,
in other words, requiring $\langle \cK\rangle=0$.

In terms of ${\cN} = 1$ superfields the formal expression of $\cK$ is
\be
\cK(x)\Big{|}_{\rm formal}  = \sum_{I=1}^{3} : 
\tr({\rm{e}}^{-2gV(x,\theta,\bar\theta)}
\Phi^{\dagger}_I(x,\theta,\bar\theta) 
{\rm{e}}^{2gV(x,\theta,\bar\theta)}
\Phi^I(x,\theta,\bar\theta)):
\Big{|}_{\theta =0, \bar \theta=0}\, ,
\label{K1naive}
\ee
where the exponents are included to ensure gauge invariance.
From~(\ref{K1naive}) it is clear that the ${\cN} = 4$ Konishi multiplet
contains the ${\cN} = 1$ Konishi submultiplet. The latter
is a real vector multiplet and among its components one finds the
classically conserved $U(1)$ axial current
\be
{\cal{K}}_{\mu} = \bar{\psi}_A \bar\sigma_{\mu} \psi^A\, .
\label{koncor}
\ee
The anomalous divergence of the Konishi current is part of the
Konishi anomaly~\cite{kenkon}, which in ${\cN} = 1$ superfield notation
reads
\be
{1\over 4} \bar{D}^{2} : \tr ({\rm e}^{-2gV}
\Phi^{\dagger}_I {\rm e}^{2gV} \Phi^I): =
\tr(\Phi^I{\partial {\cal W}\over \partial\Phi^I} ) +
{6 g^{2} N \over 32 \pi^{2}} \tr (W^{\alpha} W_{\alpha}) \, ,
\ee
where ${\cal W}$ is the superpotential, $W^{\alpha}$ is the chiral
superfield strength multiplet, defined as
\be
W_\alpha=-\frac{1}{8g}\bar{D}^2({\rm{e}}^{-2gV}D_\alpha{\rm{e}}^{2gV})
\label{walpha}
\ee
and $D$, $\bar{D}$ are the ${\cN} = 1$ supercovariant derivatives.

The presence of an anomalous divergence for the Konishi current 
(\ref{koncor}) affects the expression of the superconformal descendants of 
$\cK$. In particular at level two (\ie~after acting with two 
supersymmetry transformations), the explicit form of the
scalar operator in $\delta^{2} \cK$ reads
\be
({\cal{K}}_{\bf{10^*}})_{AB} =
3\sqrt{2}g t^{ijk}_{{AB}}\tr(\varphi_{i}\varphi_{j}\varphi_{k})
+ {6g^{2}N\over 32\pi^{2}} \tr ({\bar\psi}_{{\dot\alpha}A}
{\bar\psi}^{\dot\alpha}_{B}) \, ,
\label{k10}
\ee
where $t^{ijk}_{{AB}}$ is the totally antisymmetric product of three matrices,
$t^i_{AB}$, which in turn are the Clebsch-Gordan coefficients for the
decomposition of the vector index $i$ into two spinor indices $A,B$. The
operator $({\cal{K}}_{\bf{10^*}})_{AB}$ belongs to the representation
${\bf{10^*}}$ and bears a close resemblance to the operator 
\be
({\cal{E}}_{\bf{10^*}})_{AB} =
- \tr ({\bar\psi}_{\dot\alpha A} {\bar\psi}^{\dot\alpha}_{B})+
\sqrt{2}g t^{ijk}_{{AB}}\tr(\varphi_{i}\varphi_{j}\varphi_{k}) \, ,
\label{e10}
\ee
that is a superdescendant at level two of the chiral primary operator
\be
{\cal{Q}}^{(ij)}_{\bf 20^\prime} =
\tr \left(\varphi^i \varphi^j -{\delta^{ij} \over 6}
\varphi^k \varphi^k \right)\, .
\label{defQ}
\ee
${\cal{Q}}_{\bf 20^\prime}$ belongs to the representation ${\bf 20^\prime}$
of $SU(4)$ and, as we already said, is the lowest component of the
${\cal N}=4$ supercurrent multiplet.

The second, ``anomalous'', term in eq.~(\ref{k10}) is obviously crucial
in the construction of higher level operators and plays also a key r\^ole 
in making the two-point correlator
$\langle\delta^{2}{\cK}(x) \delta^{2}\calQ_{\bf 20^\prime}(y)\rangle$
vanish, \ie~in making ${\cal{K}}_{\bf{10^*}}=\delta^{2}\cK$ ``orthogonal'' to
${\cal{E}}_{\bf{10}} = \delta^{2}\calQ_{\bf 20^\prime}$, 
as expected on the basis of the
so-called $U_B(1)$ ``bonus symmetry''~\cite{intri}.

For later use, we need to get acquainted with the scalar operators 
of naive conformal dimension $\Delta_0=4$,
that appear at level 4 in the Konishi multiplet. In order 
to construct $\delta^{4} \cK$, it is sufficient to perform
two ${\cal N}=1$ supersymmetry transformations on
\be
({\cal{K}}_{{\bf 6^{*}}\in {\bf 10^{*}}})_{11} =
3\sqrt{2} g \, \tr(\phi_{1}^{\dagger} [\phi^2, \phi^3])
+ {6g^{2}N\over 32\pi^{2}} \tr(\bar\psi_{1}\bar\psi_{1})\label{k610} \, .
\ee
$({\cal{K}}_{{\bf 6^{*}}\in {\bf 10^{*}}})_{11} $ is one of the 
components in the ${\bf 6^{*}}$ of $SU(3)$ which appears in the 
decomposition of the ${\bf 10^{*}}$ of $SU(4)$, to which the 
operator~(\ref{k10}) belongs, with respect to the subgroup $SU(3)\otimes
U(1)$. We are confident that naive supersymmetry transformations lead
to the correct answer, since there is no need of normal-ordering the
operator $({\cal{K}}_{{\bf 6^{*}}\in {\bf 10^{*}}})_{11}$ 
that is used as a starting point.
The final result is a complex scalar operator
\bea
({\calK}_{{\bf 6^*} 
\in {\bf 20'}})_{11} &=& 6\sqrt{2} g \,
\tr([\phi_{1}^{\dagger},\psi^{2}] \psi^{3})
+ 6\sqrt{2} g^2 
\left\{ \tr([\phi_{1}^{\dagger},\phi^2][\phi^\dagger_{2},\phi_{1}^{\dagger}])
+ \tr( [\phi_{1}^{\dagger},\phi^\dagger_3][\phi^3,\phi_{1}^{\dagger}]) 
\right\} \nn \\
&+& \! {6g^{2}N\over 32\pi^{2}} \left\{ 4 \, 
\tr(D_\mu\phi_{1}^{\dagger} D^\mu\phi_{1}^{\dagger})
+ \sqrt{2} g \, \tr([\phi_{1}^{\dagger}, \bar\lambda] \bar\psi^\dagger_{1}) 
\right\}\label{calk620} \, ,
\eea
that, as indicated, belongs to the ${\bf 20'}$ of $SU(4)$. In the 
following we will refer to it as ${\calK}_{{\bf 20'}}$. 
The conjugate operator, ${\calK}^{\dagger}_{{\bf 20'}}$ enters in 
the decomposition of $\bar\delta^{4} \cK$. It is important to 
stress that neither ${\calK}_{{\bf 20'}}$ nor 
${\calK}^{\dagger}_{{\bf 20'}}$ appear in the OPE of 
${\cal{Q}}_{\bf 20^\prime}\cdot {\cK}$ at the
order we are going to work. The $U_B(1)$ symmetry~\cite{intri} would
imply this to be true at any order in $g^2$.

The definition of the superdescendants in
$\bar\delta^{2} \delta^{2} {\cK}$ is more involved. 
As far as the scalar components are
concerned, one has an operator of naive dimension $\Delta_0=4$
in the ${\bf 84}$ that can only involve scalar quadrilinear terms (terms
involving fermionic fields would have conformal dimension larger than 4)
and is consequently of the form
\be
{\cal{K}}^{ij,kl}_{\bf 84} =
g^{2}\tr \left([\varphi^{i},\varphi^{j}]
[\varphi^{k},\varphi^{l}]\right) \, .
\ee
At the same level there is an $SU(4)$ singlet of the form
\bea
{\cal K}_{\bf 1}^\prime
&=& 6 \sqrt{2} g \left\{ t^{i}_{AB}
\tr([\varphi_{i},\psi^{A}] \psi^{B}) +
t_{i}^{AB} \tr([\varphi^{i},\bar\psi_{A}] \bar\psi_{B}) \right\}
- 3 g^{2} \tr \left( [\varphi^{i},\varphi^{j}]
[\varphi_{i},\varphi_{j}] \right) \nn 
\\
&+& {6 g^2 N \over 32\pi^{2}} \left[ 2 \tr (D_\mu\varphi^i
D^\mu\varphi_i) - \tr (F^{\mu\nu}F_{\mu\nu}) \right]  \, ,
\label{k1prime}
\eea
that may be thought of as the trace of the singlet classically 
conserved symmetric tensor, ${\cal K}_{\mu\nu}$, orthogonal to the
exactly conserved and traceless canonical stress-energy tensor,
${\cal T}_{\mu\nu}$.

In addition to the components we have discussed, the ${\cal N}=4$
Konishi multiplet contains a huge number (altogether $ 2^{16}$) of other
gauge invariant composite operators. A complete classification can be
found in~\cite{AF}.

\section{Correlation functions of the Konishi operator $\cK$}

Let us consider the lowest dimensional operator, $\cK$, belonging to the
Konishi multiplet. Its gauge invariant expression is given in
eq.~(\ref{K1naive}). As we said, ${\cal K}_{\bf 1}$ has (naive)
conformal dimension $\Delta_0=2$ and a non-vanishing anomalous dimension. 
Note that, since we will only be interested in the $\theta=\bar\theta=0$
component, effectively in the exponents of eq.~(\ref{K1naive}) only the
lowest component of the vector superfield, \ie~the scalar field $c(x)$,
will be relevant. Similarly the chiral superfields 
$\Phi^I(x,\theta,\bar\theta)$
($\Phi^\dagger_I(x,\theta,\bar\theta)$) will contribute only through their
lowest components, $\phi^I(x)$ ($\phi^\dagger_I(x)$). Thus we have
\be
{\cK}(x) \Big{|}_{\rm formal}  = \sum_{I=1}^{3} : \tr({\rm{e}}^{-2gc(x)}
\phi^{\dagger}_I(x) {\rm{e}}^{2gc(x)} \phi^I(x)):\, .
\label{K1naive2}
\ee
We note that in the Fermi-Feynman gauge that we use in this paper
the propagator $\langle c(x)\,c(y) \rangle$ vanishes~\cite{kov}.

To give a precise meaning to the formal expression in eq.~(\ref{K1naive2})
one has to remember that, since the operator ${\cK}$ has an anomalous
dimension, it will suffer a non-trivial renormalization. We find it
convenient to regularize ${\cK}$ as suggested by the OPE, \ie~by point 
splitting
\be
{\cK}(x)\Big{|}_{\rm reg}   =  a(g^2)\sum_{I=1}^{3}:\tr({\rm{e}}^{-2gc(x)}
\phi^{\dagger}_I(x+{\epsilon \over
2}){\rm{e}}^{2gc(x)}\phi^I(x-{\epsilon \over 2})):\, ,
\label{K1ansatz} 
\ee 
where $\epsilon$ is an infinitesimal, but otherwise arbitrary, 
four-vector. $a(g^2)$ is a normalization factor that we 
choose of the form $a(g^2)=1+g^2 a_1+g^4 a_2 +\ldots $. Unlike 
the operators corresponding to symmetry generators (like the R-symmetry 
currents or the stress-energy tensor), the Konishi scalar $\cK$ has 
no intrinsic normalization, so we shall use this freedom in the following to 
conveniently fix its normalization. 

Note that, owing to our choice of gauge-fixing, there is no need
to ``point-split'' the vector field in the exponents, because, as
observed above, the $c$-field has vanishing propagator. An
additional refinement to the formula~(\ref{K1ansatz}) could be to
make the replacement
\be
{\rm{e}}^{gc(x)}\rightarrow {\rm{e}}^{g\int_{-1}^1 d\rho
c(x+\rho{\epsilon \over 2})}\, , \label{wilsonline} 
\ee 
but this would not affect the anomalous dimension of the operator and would 
only lead to a finite rescaling, which can be compensated by an appropriate 
change of the normalization constant, $a(g^2)$ in~(\ref{K1ansatz}). So in 
the following we shall stick to the simpler regularized 
expression~(\ref{K1ansatz}).

By conformal invariance, renormalization of $\cK$ simply amounts to a 
multiplication by the ``renormalization constant'' 
\be
Z(\mu\epsilon)=(\mu^2\epsilon^2)^{-\gamma_{\cK}(g^2)/2}\, ,
\label{renconst}
\ee
where $\gamma_{\cK}(g^2)$ is the anomalous dimension of $\cK$ 
(by superconformal invariance $\gamma_{\cK}(g^2)$ is the anomalous dimension 
of the whole Konishi supermultiplet). We assume (as is always the case 
in perturbation theory) that $\gamma_{\cK}(g^2)$ is small and represented by 
the series expansion
\be 
\gamma_{\cK}(g^2)=
g^2\gamma_1+g^4\gamma_2+\ldots\label{gammak1} \label{gammak} \, . 
\ee
{}From the results of refs.~\cite{ans} and~\cite{bkrs1,bkrs2} the first
two coefficients of the expansion are known and (after the
standard definition of $g$ is used) are given by
\ba
&&\gamma_1=\frac{3N}{4\pi^2}\label{gammak11}\\
&&\gamma_2=-\frac{3N^{2}}{16\pi^4}\label{gammak12}\, .
\ea

Below we show that two-, three- and four-point Green functions of $\cK$ are
made finite (up to order we have computed them), if the renormalized 
operator is taken to be of the form
\be
{\cK}(x)\Big{|}_{\rm renorm}   =  {a(g^2) \over (\epsilon^2)^{{1 \over
2}\gamma_{\cK}(g^2)}}\sum_{I=1}^{3}:\tr({\rm{e}}^{-2gc(x)}
\phi^{\dagger}_I(x+{\epsilon \over 2}){\rm{e}}^{2gc(x)}
\phi^I(x-{\epsilon \over 2}):\, ,
\label{K1ren} \ee 
provided the expansion~(\ref{gammak}) is used with $\gamma_1$ and $\gamma_2$ 
as in eqs.~(\ref{gammak11}) and~(\ref{gammak12}). In this equation and 
in the following, to avoid cumbersome formulae, we
refrain from displaying explicitly the obvious $\mu$ dependence.

As a check of the correctness of eq.~(\ref{K1ren}), one can
compute the O($g^2$) corrections to the two- and three-point
functions of ${\cK}$. The coordinate dependence of these correlators
is completely determined by conformal invariance. We shall use the
freedom in the normalization factor $a(g^2)$ in eq.~(\ref{K1ren}),
to make the normalization of the two-point function independent of $g^2$.
This is achieved at the order we work by setting $a_1 = {3 N /8 \pi^2 }$
(the other coefficients would require higher order computations to be fixed). 
With this choice one finds for the two-point function
\be \langle {\cK}(x_1){\cK}(x_2)\rangle = {3(N^2-1) \over (4 \pi^2)^2}
{1 \over({x_{12}}^2)^{2+\gamma_{\cK}(g^2)}} \, , 
\label{2pK1} 
\ee
and for the three-point function
\be \langle
{\cK}(x_1){\cK}(x_2){\cK}(x_3)\rangle = {c_{\calK\calK\calK}(g^2)
\over ({x_{12}}^2{x_{13}}^2{x_{23}}^2)^{1+{1 \over
2}\gamma_{\cK}(g^2)}}\, , 
\label{3pK1} 
\ee
where the cubic coupling is given by
\be
c_{\calK\calK\calK} (g^2)= \frac{3(N^2-1)}{4(4\pi^2)^3} + O(g^2)\, .
\label {ckkk}\ee
Taking $\gamma_1={3N/ 4\pi^2}$, as in eq.~(\ref{gammak}), the
$g^2$-expansion of the above expressions turns out to be in complete
agreement with the results of the perturbative calculations, which give
\be \langle {\cK}(x_1){\cK}(x_2)\rangle
\Big{|}_{g^2}=-{9N(N^2-1) \over 4(4\pi^2)^3}{1\over x_{12}^4}
\ln(x_{12}^2) \ , \label{2pK1g2} \ee \be \langle
{\cK}(x_1){\cK}(x_2){\cK}(x_3)\rangle \Big{|}_{g^2} = - {9 N
(N^2-1) \over 8(4 \pi^2)^4}{1 \over x_{12}^2 x_{13}^2 x_{23}^2}
[\ln(x_{12}^2 x_{13}^2 x_{23}^2) + 3]  \label{3pK1g2} \ee
for the O($g^2$) corrections to 
two- and three-point correlators, respectively.

The tree-level value of the Green function of four Konishi scalars is
\ba
\langle {\cK}(x_1){\cK}(x_2){\cK}(x_3){\cK}(x_4)\rangle &&
\!\!\!\!\!\!\!\!\!\!\!\!\!
 \Big{|}_{\rm tree} = { (N^2-1) \over 16 (4
\pi^2)^4 } \left[ {9 (N^2-1) \over x_{12}^4 x_{34}^4 } + {9
(N^2-1) \over x_{14}^4 x_{23}^4 } + {9 (N^2-1) \over x_{13}^4
x_{24}^4 } \right.
\nonumber \\
 &+&  \left. {6 \over x_{12}^2 x_{34}^2 x_{13}^2 x_{24}^2 } + {6
\over x_{12}^2 x_{34}^2 x_{14}^2 x_{23}^2 } + {6 \over x_{14}^2
x_{23}^2 x_{13}^2 x_{24}^2 }  \right]\, , \label{K4tree} \ea
while for the O($g^2$) correction we find
\ba
&&\langle {\cK}(x_1){\cK}(x_2){\cK}(x_3){\cK}(x_4)\rangle
\Big{|}_{g^2}=\nonumber \\
&&- {3 N(N^2-1) \over 16 (4 \pi^2)^5 }
\left[ {B(r,s) \over x_{12}^2 x_{34}^2 x_{14}^2 x_{23}^2}
(1+r^2+s^2+4r+4s+4rs) \right. \nonumber  \\
&&+ {6 \over x_{12}^2 x_{34}^2 x_{13}^2 x_{24}^2 } \left(\ln
\left( { x_{12}^2 x_{34}^2 x_{13}^2 x_{24}^2 \over x_{14}^2
x_{23}^2 } \right) +2 \right) + {6 \over x_{12}^2 x_{34}^2
x_{14}^2 x_{23}^2 } \left(\ln \left( { x_{12}^2 x_{34}^2 x_{14}^2
x_{23}^2 \over x_{13}^2
x_{24}^2 } \right) +2 \right) \nonumber \\
&&+ {6 \over x_{14}^2 x_{23}^2 x_{13}^2 x_{24}^2 } \left(\ln
\left( { x_{14}^2 x_{23}^2 x_{13}^2 x_{24}^2 \over x_{12}^2
x_{34}^2 } \right) +2 \right) + {9 (N^2-1) \over x_{12}^4 x_{34}^4
}
\ln \left( { x_{12}^2 x_{34}^2} \right) \nonumber \\
&&+ \left. {9 (N^2-1) \over x_{14}^4 x_{23}^4 } \ln \left( {
x_{14}^2 x_{23}^2} \right)  + {9 (N^2-1) \over x_{13}^4 x_{24}^4 }
\ln \left( { x_{13}^2 x_{24}^2} \right)   \right]\, , \label{KKKK}\ea
where the massless scalar box integral
\ba B(r,s) & = & {1 \over \sqrt{p}} \left \{\ln (r)\ln
(s)  - \left [\ln \left({r+s-1 -\sqrt {p} \over 2}\right)
\right]^{2} \right. \nonumber \\
&& \left. -2 {\rm Li}_2 \left({2 \over 1+r-s+\sqrt {p}}\right ) -
2 {\rm Li}_2 \left({2 \over 1-r+s+\sqrt {p}}\right )\right \}\, ,  
\label{Brsf}
\ea
is a function of the two conformally invariant ratios
\be
r = { x_{12}^2 x_{34}^2 \over x_{13}^2 x_{24}^2 } \quad  , \qquad
s ={x_{14}^2 x_{23}^2\over x_{13}^2 x_{24}^2} \, .
\label{rs}
\ee
In eq.~(\ref{Brsf}) we have introduced the definition
\be
p = 1 + r^{2} + s^{2} - 2r - 2s - 2rs \, .
\label{pdef}
\ee
and assumed proper analytic continuation of 
${\rm Li}_2 (z)=\sum_{n=1}^{\infty}{z^n\over n^2}$.

Note that, contrary to the case of the four-point functions involving 
only protected operators, which depend only on the conformally invariant 
cross ratios~(\ref{rs}), the four-point function~(\ref{KKKK}) has an 
explicit logarithmic dependence on coordinate differences. This 
behaviour, which is similar to what one finds for the two- and three-point
functions~(\ref{2pK1g2}) and~(\ref{3pK1g2}), does not contradict
conformal invariance and it is just a manifestation of
the anomalous dimension of the Konishi scalar ${\cK}$.

Let us now consider Green functions involving both the protected operators
${\cal{Q}}_{\bf 20^\prime}$ in (\ref{defQ}) and the unprotected operator
${\cK}$.

We recall that the case of the four protected operators
${\cal{Q}}_{\bf 20^\prime}^{(ij)}$ has
been studied previously. The relevant correlators are known explicitly
up to O($g^4$) and at the one-instanton level~\cite{bkrs1,bkrs2,bgkr,ess}.

In terms of $SU(3)\otimes U(1)$ the ${\cal Q}_{\bf 20^\prime}^{(ij)}$'s
decompose in
\be
{\cal{C}}^{IJ}(x)=\tr(\phi^I(x)\phi^J(x)) \, ,
\qquad {\cal{C}}^{\dagger}_{IJ} (x) =
\tr(\phi^{\dagger}_I (x) \phi^{\dagger}_J (x))
\ee
and
\be
{\cal{V}}^I_J = \tr\left({\rm{e}}^{-2gc(x)}
\phi^{\dagger}_J(x){\rm{e}}^{2gc(x)}\phi^I(x)\right)-
{\delta^I_J\over 3}\tr\left({\rm{e}}^{-2gc(x)} \phi^{\dagger}_L(x)
{\rm{e}}^{2gc(x)} \phi^L(x)\right) \, ,
\ee
where again the operators have been regularized
by point-splitting like in eq.~(\ref{K1ansatz}).
Note that no normal-ordering is
needed as the {\it vev}'s of all these operators vanish,
since they are not $SU(4)$ singlets.

As an additional check of the correctness of our approach,
we may compute the three-point function of two protected operators
$\calQ_{\bf 20^\prime}$ and one ${\cK}$, for which again we find a
perturbative expression in perfect agreement with the form
required by conformal invariance, which reads
\be \langle
{\cal{Q}}_{\bf 20^\prime}^{(i_1 j_1)}(x_1) {\cal{Q}}_{\bf
20^\prime}^{(i_2 j_2)}(x_2) {\cK}(x_3)\rangle =
{{c_{\calQ^{(i_1,j_1)} \calQ^{(i_2 j_2)} \calK}(g^2)}
\over {{x_{12}}^2{x_{13}}^2{x_{23}}^2}}
\left({{x_{12}}^2 \over {x_{13}}^2{x_{23}}^2} \right)^{{1 \over
2}\gamma_{\cK}(g^2)} \, . \ee
Indeed one finds at tree-level
 \be \langle
{\cal{C}}^{11}(x_1){\cal{C}}^{\dagger}_{11}(x_2) {\cK}(x_3)\rangle
\Big{|}_{\rm tree} = { (N^2-1) \over 2 (4 \pi^2)^3 } {1 \over
x_{12}^2 x_{13}^2 x_{23}^2 } \label{CCdKtree} \ee
and at O($g^2$)
 \be \langle
{\cal{C}}^{11}(x_1){\cal{C}}^{\dagger}_{11}(x_2) {\cK}(x_3)\rangle
\Big{|}_{g^2} = { 3 N(N^2-1) \over 16 (4 \pi^2)^4 } {1 \over
x_{12}^2 x_{13}^2 x_{23}^2 }
 \left( \ln
\left( {x_{12}^2 \over x_{13}^2 x_{23}^2} \right) -1 \right) \, .
\label{CCdKg2} \ee
Again this result is consistent with the known
O($g^2$) value of the $\cK$ anomalous dimension.

Passing to four-point Green functions, there are only two
non-vanishing choices \be \langle {\cal{Q}}_{\bf
20^\prime}^{(i_1 j_1)}(x_1) {\cal{Q}}_{\bf 20^\prime}^{(i_2
j_2)}(x_2) {\cK}(x_3) {\cK}(x_4)\rangle \label{QQKK} \ee \be
\langle{\cal{Q}}_{\bf 20^\prime}^{(i_1j_1)}(x_1) {\cal{Q}}_{\bf
20^\prime}^{(i_2j_2)}(x_2) {\cal{Q}}_{\bf
20^\prime}^{(i_3j_3)}(x_3) {\cK}(x_4)\rangle \, , \label{QQQK} \ee
because the expectation value of three ${\cK}$ with one
${\cal{Q}}_{\bf 20^\prime}^{(ij)}$ is trivially zero due to
$SU(4)$ symmetry.

To avoid cumbersome notations, instead of writing correlators for
generic $SU(4)$ labels, we shall choose representatives. Note that 
reconstructing the general expression of the amplitudes is immediate,
since in the cases of interest there is only one independent $SU(4)$
structure for both the Green functions~(\ref{QQKK}) and~(\ref{QQQK}). 
In fact in the corresponding $SU(4)$ tensor product there
is only one singlet, unlike what happens for the $\Q20\Q20\Q20\Q20$
product, where there are six singlets (related by permutations and by a 
non-trivial functional relation~\cite{bkrs2,ess}).
In particular we find for the Green function with two
${\cal{Q}}_{\bf 20^\prime}$'s and two ${\cK}$'s at tree-level
\ba 
&&\langle {\cal{C}}^{11}(x_1){\cal{C}}^{\dagger}_{11}(x_2)
{\cK}(x_3){\cK}(x_4)\rangle
\Big{|}_{\rm tree} =  \nonumber \\
&& { (N^2-1) \over 4 (4 \pi^2)^4 } \left[ {1 \over x_{12}^2
x_{34}^2 x_{13}^2 x_{24}^2 }  + {1 \over x_{12}^2 x_{34}^2
x_{14}^2 x_{23}^2 } + {1 \over x_{14}^2 x_{23}^2 x_{13}^2 x_{24}^2}
 + { 3 (N^2-1) \over 2 x_{12}^4
x_{34}^4 }   \right] \, , \label{CCdKKtree} \ea
while the O($g^2$) correction is
\ba
&&\langle {\cal{C}}^{11}(x_1){\cal{C}}^{\dagger}_{11}(x_2)
{\cK}(x_3){\cK}(x_4)\rangle
\Big{|}_{g^2} =  \nonumber \\
&&- { N(N^2-1) \over 8 (4 \pi^2)^5 }
\left[ {B(r,s) \over x_{12}^2 x_{34}^2 x_{14}^2 x_{23}^2}
(1+r^2+s^2-2r+4s-2rs) \right. \nonumber  \\
&&+ {3 \over x_{12}^2 x_{34}^2 x_{13}^2 x_{24}^2 } \left(\ln
\left( { x_{34}^4 x_{13}^2 x_{24}^2 \over x_{14}^2 x_{23}^2 }
\right) +2 \right) + {3 \over x_{12}^2 x_{34}^2 x_{14}^2 x_{23}^2
} \left(\ln \left( { x_{34}^4 x_{14}^2 x_{23}^2 \over x_{13}^2
x_{24}^2} \right) +2 \right) \nonumber \\
&&+ \left. {3 \over x_{14}^2 x_{23}^2 x_{13}^2 x_{24}^2 }
\left(\ln \left( { x_{14}^2 x_{23}^2 x_{13}^2 x_{24}^2 \over
x_{12}^4 } \right) +2 \right) + { 9 (N^2-1) \over x_{12}^4
x_{34}^4 }  \ln \left( { x_{34}^2} \right)   \right] \, .
\label{CCdKK} 
\ea
This function is of particular interest, since in the 3-4-- and 
2-3--channels ($x_{34}\rightarrow 0$ and $x_{23}\rightarrow 0$, 
respectively) it gets contributions from operators in well defined 
$SU(4)$ representations. Indeed only the singlet can contribute in the
3-4--channel and only the representation ${\bf 20'}$ in the
2-3--channel.

For the Green function with three ${\calQ}_{\bf 20^\prime}$'s and one
${\cK}$, we find at tree-level
\ba
&&\langle {\cal{C}}^{11}(x_1) {\cal{C}}^{\dagger}_{11}(x_2)
{\cal{V}}_2^2(x_3) {\cK}(x_4)\rangle
\Big{|}_{\rm tree} =  \nonumber \\
&&- { (N^2-1) \over 12 (4 \pi^2)^4 } \left[ {1 \over x_{12}^2
x_{34}^2 x_{13}^2 x_{24}^2 }  + {1 \over x_{12}^2 x_{34}^2
x_{14}^2 x_{23}^2 } + {1 \over x_{14}^2 x_{23}^2 x_{13}^2 
x_{24}^2} \right] 
\label{CCdVKtree} \ea
and at O($g^2$)
\ba 
&&\langle {\cal{C}}^{11}(x_1)
{\cal{C}}^{\dagger}_{11}(x_2) {\cal{V}}_2^2(x_3) {\cK}(x_4)\rangle
\Big{|}_{g^2} =  \nonumber \\
&& { N(N^2-1) \over 24 (4 \pi^2)^5 }
\left[ {B(r,s) \over x_{12}^2 x_{34}^2 x_{14}^2 x_{23}^2}
(1+r^2+s^2-2r-2s-2rs) \right. \nonumber  \\
&&- {3 \over x_{12}^2 x_{34}^2 x_{13}^2 x_{24}^2 } \left(\ln
\left( { x_{23}^2 \over x_{24}^2 x_{34}^2 } \right) - 1 \right) -
{3 \over x_{12}^2 x_{34}^2 x_{14}^2 x_{23}^2 } \left(\ln \left( {
x_{13}^2  \over x_{14}^2 x_{34}^2 } \right) - 1
\right) \nonumber \\
&&- \left. {3 \over x_{14}^2 x_{23}^2 x_{13}^2 x_{24}^2 }
\left(\ln \left( { x_{12}^2  \over x_{14}^2 x_{24}^2 } \right) - 1
\right) \right] \, . \label{CCdVK} \ea
In this case in all channels only the representation ${\bf 20'}$
can be exchanged.

\section{Anomalous dimensions and OPE}

In this section we analyze the general structure of the consistency 
conditions that can be obtained combining conformal invariance, which
requires power-like expressions for the Green functions, with
the observed logarithmic behaviour found in perturbation theory. 
In particular an exact resummation of the logarithms to all orders in $g^2$
is expected to take place. Although this has been explicitly checked so 
far only at order O($g^4$) for the lowest operator of the supercurrent 
multiplet, we shall assume it to be true in general, since otherwise
scale invariance would be violated.

We shall consider a general four-point function
of not necessarily protected operators. 
We schematically  write it in the
form $\langle Q_1(x_1)Q_2(x_2)Q_3(x_3)Q_4(x_4) \rangle$, and work in the 
double OPE limit, say, $x_1 \rightarrow x_2$ and simultaneously $x_3 
\rightarrow x_4$ ($s$-channel). In this limit the perturbative 
corrections to the four-point functions exhibit a logarithmic 
behaviour, which, as we said, is a manifestation of the non-vanishing 
anomalous dimensions of the external operators and/or of the operators 
exchanged in the intermediate channel. For simplicity we shall 
concentrate only on the leading logarithmic contributions, that 
behave like $\ln^n(x_{ij}^2)$ at order O($g^{2n}$), but similar 
considerations are valid also for subleading logarithmic terms.

Let us consider the logarithmic contributions coming from a finite 
set of operators $O_i$, $i=1 \dots k$, all with the same tree-level scale 
dimension, ${\Delta}_{0}$, but with different O($g^2$) anomalous dimensions.
Note that operators with the same anomalous dimension cannot be separated 
unambiguously in this approach. We assume that the intermediate operators 
have been made orthogonal at tree-level in the sense that they are 
chosen to satisfy the equations
\be \langle O_i(x_i) O_j(x_j)\rangle\Big{|}_{\rm tree} =
\delta_{ij} { N_i \over (x_{ij}^2)^ {{\Delta}_{0}}}\, ,
\label{ort} \ee 
with $N_i>0$ due to positivity.

The coefficients of the leading logarithm $\ln^n (x_{24}^2)$ 
(due to conformal invariance, consideration of the other two leading 
logarithmic behaviours, \ie~$\ln^n (x_{12}^2)$ and $\ln^n (x_{34}^2)$ 
does not lead to independent conditions) will satisfy the following relations
\begin{itemize}
\item tree-level (corresponding to $n=0$)
 \be \sum_{i=1}^k F_i = P_0 \,  , \label{tree} \ee
\item order O($g^2$) (corresponding to $n=1$)
 \be \sum_{i=1}^k F_i \gamma_i = P_1 \,  , \label{oneloop} \ee
\item order O($g^4$) (corresponding to $n=2$)
 \be \sum_{i=1}^k F_i (\gamma_i)^2= P_2 \, , \label{twoloop}
\ee 
\end{itemize}
where $\gamma_i$ is the O($g^2$) correction to the anomalous dimension 
of the $i$-th operator, $O_i$, and the coefficients, $F_i$, are ratios
of tree-level normalization constants 
\be F_i=  { c_{ Q_1 Q_2 O_i} c_{Q_3 Q_4 O_i} 
 \over N_i  } \, . \label{F's}
 \ee
$P_n$ is what results from the explicit perturbative
calculations, after the contributions of the leading
operators have been removed. Here we shall sketch the procedure 
one has to follow in order to compute $P_n$ in the case of interest 
(scalars of naive dimension $\Delta_0=4$). The generalisations to 
higher $\Delta_0$ and to tensor operators is straightforward, 
though algebraically rather involved. One first expands in double 
power series for small $x_{12}$ and $x_{34}$ the correlator
\be
(-1)^n n! \langle Q_1(x_1) Q_2(x_2) Q_3(x_3) Q_4(x_4) \rangle
\Big{|}_{g^{2n}} \, ,
\ee  
keeping only the terms proportional to $\ln^n (x_{24}^2)$.
As already noted, consideration of the other leading logarithmic terms, 
proportional to $\ln^n (x_{12}^2)$ and $\ln^n (x_{34}^2)$, would lead
to equivalent conclusions, but the equations coming from the 
$\ln^n (x_{24}^2)$ terms are simpler, since they are manifestly
independent of the anomalous dimensions of the external operators.
The result of the expansion has (in general) power singularities
for small $x_{12}$ and $x_{34}$. These come from intermediate
scalar operators of naive dimension $\Delta_0^S=2$, from vectors 
of naive dimension $\Delta_0^V=3$ and from symmetric rank two 
traceless tensors of naive dimension $\Delta_0^T=4$. In order to 
single out the contribution of the scalars of 
dimension $\Delta_0=4$ we are interested in, one has to subtract 
the contributions of all the above operators, together with their 
derivatives (descendants). This procedure is made possible by
the fact that the coefficients with which 
descendants contribute are completely determined by conformal 
invariance. Notice that contributions coming from the second 
derivative of a scalar, as well as those coming from 
the first derivative of a vector or the trace of a tensor
give rise to regular behaviours. Consequently, their subtraction 
is crucial to get the correct value of $P_n$, which is the residual 
coefficient of the regular term we are after. 

The generalization of eqs.~(\ref{tree}) to~(\ref{twoloop}) to 
order O($g^{2n}$) is straightforward and reads
\be \sum_{i=1}^{k} F_i(N) (\gamma_i(N))^n = P_n(N) \,  , \label{Pnloop}
\ee 
where we made explicit the dependence of both $F_i$ and $P_n$
on the number of colours $N$. Note that the $N$ dependence of
$P_n$ for $n \leq 3$ is polynomial for all four-point functions we
computed. Indeed the $N$ dependence of the four-point functions
comes only from the traces over the colour indices which can be
rewritten as traces of products of an even number of $SU(N)$ matrices
in the adjoint representation. For $\ell=2,4,6$ matrices only
the ``planar'' factor $(N)^{\ell} (N^{2} -1)$ appears. 
For more than eight matrices, non-planar contributions start to 
appear with colour factors that will depend on the exact structure of the 
four-point function under consideration. 

Eliminating the quantities $F_i(N)$ from the system~(\ref{Pnloop}) 
leads to the following consistency equations
\ba 
&& P_{k+L} -(\sum_i \gamma_i)
P_{k+L-1} + (\sum_{i<j} \gamma_i \gamma_j) P_{k+L-2}
-(\sum_{i<j<l} \gamma_i \gamma_j \gamma_l) P_{k+L-3} + \ldots \nonumber\\
&& + (-1)^k \gamma_1 \gamma_2 \ldots \gamma_k P_L =0
\label{neweqs} \ea 
for any $L\ge 0$. Eqs.~(\ref{neweqs}) imply that  the
combinations of anomalous dimensions
\be
\sum_i \gamma_i \, , \qquad 
\sum_{i<j} \gamma_i \gamma_j \, ,  \qquad
\sum_{i<j<l} \gamma_i \gamma_j \gamma_l \, ,  \qquad \dots  \qquad
\gamma_1 \gamma_2 \ldots \gamma_k  
\label{comb} 
\ee 
as well as all the $P_n$'s are
completely determined, once one knows the leading logarithmic behaviour
up to order O($(g^2)^{2k-1}$). In fact to solve for the $k$ 
variables~(\ref{comb}) one needs eq.~(\ref{neweqs}) for $k$ 
different values of $L$. Thus the knowledge of the coefficients
$P_n$ from $n=0$ to $n=2k-1$ is enough to determine everything. 
Unfortunately, except for some special cases, the currently 
available data is far from meeting the minimal information 
required, even when $k$ is relatively small (\eg~$k=3$ or $k=4$). 
Considering several different correlators only partially improves 
the situation, since the number of unknowns is anyway very large.

On the other hand the  knowledge of only $P_0,P_1$ and $P_2$
allows one to obtain bounds for the smallest and largest anomalous
dimensions of the operators contributing to the four-point functions
of two identical pairs of operators of the form $\langle Q_1 Q_2 Q_1 Q_2
\rangle$, even without knowing the explicit expressions of the 
intermediate operators $O_i$.

Indeed if $\gamma_{\rm min}$ is the smallest anomalous dimension (or any
of them if there are several), then multiplying eq.~(\ref{tree}) by
$\gamma_{\rm min}$ and subtracting it from eq.~(\ref{oneloop}), one obtains
\be 
\sum_i F_i (\gamma_i - \gamma_{\rm min}) = (P_1 - \gamma_{\rm min} P_0) \,.
\label{1limp} 
\ee
The left hand side is non-negative, since by
assumption $\gamma_{\rm min}$ is smaller than or equal to all the other
$\gamma_i$, while $F_i$ are non-negative due to the assumption $Q_3
= Q_1$, $Q_4 = Q_2$. One thus derives the inequality 
\be 
\gamma_{\rm min} \leq {P_1 \over P_0} \, . 
\label{P1/P0small} 
\ee
Calling $\gamma_{\rm Max}$ the largest of the anomalous dimension 
and repeating the previous argument, we find
\be
\gamma_{\rm Max} \geq {P_1 \over P_0} \, . 
\label{P1/P0large} 
\ee
Both in (\ref{P1/P0small}) and in (\ref{P1/P0large}) the equality 
is reached if all the operators have the same
anomalous dimension.

In the same way, if $\gamma_{\rm s}$ is the
anomalous dimension with smallest square (or any of them if there
are several), then multiplying eq.~(\ref{tree}) by $\gamma_{\rm s}^2$
and subtracting it from eq.~(\ref{twoloop}), one gets
\be \sum_i
F_i((\gamma_i)^2-(\gamma_{\rm s})^2)=(P_2-(\gamma_{\rm s})^2 P_0)\, ,
\label{2limp} \ee
implying
 \be  \gamma_{\rm s}^2 \leq {P_2 \over P_0 }\, . \label{P2/P0}
\ee

An important implication of these considerations is that, if $P_2=0$ 
(which implies also $P_1=0$) in some Green function, then the O($g^2$) 
corrections to the anomalous dimensions of all the operators that 
contribute to it are zero, since all the non-negative products
$F_i \gamma_i^2$ vanish. Thus for each $i$ one either has $F_i=0$, 
which means that the corresponding operator is not present in the 
OPE, or $\gamma_i=0$. Let us stress that the vanishing of $P_1$ alone 
cannot guarantee the vanishing of the anomalous dimensions, due to 
possible cancellations among positive and negative terms. However, 
if $P_1$ is negative, then it is immediate to deduce that at least one 
of the anomalous dimensions (and in particular the smallest one) 
must be negative.

\section{The {\bf icosaplet} and the singlet}

In this section we will present the results one can get for the
anomalous dimensions of the operators of naive conformal dimension
$\Delta_0=4$ in the singlet and in the icosaplet
(the representation ${\bf 20'}$), making use of the O($g^2$)
computations reported in the previous section and of the O($g^4$)
results obtained in ref.~\cite{bkrs2,aeps}.

For the purpose of isolating the representations of interest it
turns out to be convenient to consider the correlators 
\be 
\langle [{\cal{C}}^{11}(x_1) {\cal{C}}^{\dagger}_{11}(x_2)-
{\cal{C}}^{22}(x_1) {\cal{C}}^{\dagger}_{22}(x_2)]
[{\cal{C}}^{11}(x_3) {\cal{C}}^{\dagger}_{11}(x_4) -
{\cal{C}}^{22}(x_3) {\cal{C}}^{\dagger}_{22}(x_4)]\rangle
\label{QQQQ20} \ee 
for the study of the icosaplet and 
\be
\sum_{I,J=1}^{3}\langle {\cal{C}}^{II}(x_1)
{\cal{C}}^{\dagger}_{II}(x_2) {\cal{C}}^{JJ}(x_3)
{\cal{C}}^{\dagger}_{JJ}(x_4)\rangle \label{QQQQ1} \ee 
for the singlet. Both the above correlators in the relevant channels
actually receive contributions also from $\Delta_0=4$ operators
belonging to other representations, namely the ${\bf 105}$ and the
${\bf 84}$. While all the operators in the ${\bf 105}$ are
protected~\footnote{The representation ${\bf 105}$ has Dynkin
labels $[0,4,0]$.}, there exist operators in the ${\bf 84}$ that
belong to the unprotected Konishi supermultiplet. Hence a
laborious subtraction procedure is required to single out the
contribution of individual representations.

Let us start discussing the case of the representation 
${\bf 20'}$. There are 6 possible operators of naive dimension 
$\Delta_0 = 4$. Two of them, ${{\cal{K}}_{\bf 20^\prime}}$ and 
${{\cal{K}}_{\bf 20^\prime}}^\dagger$, are of the Yukawa type
at leading order in $g$ (see eq.~(\ref{calk620})) and do not contribute 
to functions of only scalars at the order we work. The other four 
operators are at leading order purely scalar (since they have to be 
orthogonal to ${{\cal{K}}_{\bf 20^\prime}}$ and 
${{\cal{K}}_{\bf 20^\prime}}^\dagger$). In preparation to our later 
analysis, it is convenient to split them into a double trace operator
\be
D^{(ij)}_{{\bf 20'}}(x) = :{\cal{Q}}_{\bf 20^\prime}^{(ik)}(x)
{\cal{Q}}_{\bf 20^\prime}^{(jk)}(x)- {\delta_{ij} \over
6}{\cal{Q}}_{\bf 20^\prime}^{(kl)}(x) {\cal{Q}}_{\bf
20^\prime}^{(kl)}(x): \label{D20}\ee
which appears in the OPE of ${\cal{Q}}_{\bf 20^\prime}^{(i_1 j_1)}(x)
\cdot {\cal{Q}}_{\bf 20^\prime}^{(i_2 j_2)}(y)$ and three addittional 
operators, 
spanning a 3-dimensional space orthogonal to $D_{{\bf 20'}}$. In this
space it is convenient to take as reference ``directions'' the 
double trace operator 
\be \hat{M}^{(ij)}_{{\bf 20'}}(x) =
:{\cal{Q}}_{\bf 20^\prime}^{(ij)}(x){\cK}(x):\,-{6 \over 3 N^2-2}
D^{(ij)}_{{\bf 20'}}(x)\, , \label{Mhat20}\ee 
which couples to ${\cal{Q}}_{\bf 20^\prime}^{(ij)}(x)\cdot 
{\cK}(y)$ and two single trace operators that vanish for $SU(2)$. 
One, $L_{{\bf 20'}}$, is proportional to the quartic Casimir of the 
$SU(N)$ gauge group, the other, $O_{{\bf 20'}}$, is a
linear combination of $D_{{\bf 20'}}$, $\hat M_{{\bf 20'}}$ and 
$\tr([\varphi^i,\varphi^k] [\varphi^k,\varphi^j]) - { \delta^{ij} \over
6} \tr( [\varphi^l,\varphi^k] [\varphi^k,\varphi^l])$.

After subtracting out the contribution of the lowest dimensional 
operators, we find from eq.~(\ref{QQQQ20}) $P_2 =0$. According
to our previous discussion, this means that the O($g^2$) correction 
to the anomalous dimension of $D_{{\bf 20'}}$, which we know has
a non-zero coupling to ${\cal{Q}}_{\bf 20^\prime}
\cdot {\cal{Q}}_{\bf 20^\prime}$, has to vanish. The three other 
scalar operators are orthogonal to $D_{{\bf 20'}}$. 
Tree-level orthogonality (eq.~(\ref{ort})) is enough to ensure the
absence of contributions from these operators to all 
logarithmic corrections (even to the subleading ones) at order O($g^4$). 
Consequently, in agreement
with~\cite{aeps}, we conclude that also the O($g^4$) correction 
to the anomalous dimension of $D_{{\bf 20'}}$ is zero. The 
vanishing of the O($g^2$) anomalous dimension was first pointed out
in~\cite{afp}. The absence of O($g^4$) correction confirms the
conjecture made by these authors that $D_{{\bf 20'}}$ is
protected, although there is no known shortening condition
associated to its quantum numbers. 

Since the analysis of the amplitude $\langle{\cal C}{\cal C}^{\dagger}
{\cal V}{\calK}_{\bf1}\rangle$ does not lead 
to independent relations (it merely confirms previous O($g^2$) results), 
we are left with only one constraint from the Green function 
$\langle{\cal C}{\cal C}^{\dagger}{\cal K}_{\bf 1}{\calK}_{\bf1} 
\rangle$ (eqs.~(\ref{CCdKKtree}) and~(\ref{CCdKK})). 
Although this is not enough to determine the 
six remaining unknowns (the 3 angles of 
mixing and the 3 anomalous dimensions of the operators 
$\hat M_{{\bf 20'}}$, $O_{{\bf 20'}}$ and $L_{{\bf 20'}}$), we can, 
nevertheless, obtain bounds for the order O($g^2$) corrections to the 
anomalous dimensions of the operators appearing in the OPE of
${\cal{Q}}_{\bf 20^\prime}^{(i j)}(x)\cdot {\cK}(y)$, namely
\be 
\gamma_{\rm min} \leq  
{(3 N^2-2) \over (N^2-2)} \times  {g^2 N \over 4 \pi^2}  
\leq \gamma_{\rm Max}  \, . 
\label{inequal}
\ee 
{}From eq.~(\ref{inequal}) we conclude that in the limit 
$N \rightarrow \infty$, $g^2N$ fixed, $\gamma_{\rm Max}$ 
will be non-vanishing, like what happens for $\cK$.

The case of $SU(2)$ is peculiar, since both $O_{{\bf 20'}}$ 
and $L_{{\bf 20'}}$ vanish, hence the only relevant operator 
is $\hat M_{{\bf20'}}$. Its anomalous dimension can then be 
determined and turns out to be equal to $5\times{2g^2\over 4\pi^2}$.

In the singlet sector again there are too many operators and from 
the computations we have at our disposal we can only get bounds 
for the relevant anomalous dimensions. 
The smallest possible value for the O($g^2$) anomalous dimension 
of the $\Delta_0=4$ singlet operators contributing to the OPE of 
${\cal{Q}}^{(i_1 j_1)}_{\bf 20^\prime}(x) \cdot 
{\cal{Q}}_{\bf 20^\prime}^{(i_2 j_2)}(y)$ 
turns out to be negative and to satisfy the bound 
\be 
\gamma_{\rm min} \leq  -{12 \over(3N^2-1)} \times 
{g^2 N \over 4 \pi^2}  \leq 0 \, . 
\ee
For the smallest square, $\gamma_{\rm s}$ ($\gamma_{\rm s}$ may be 
different from $\gamma_{\rm min}$), one finds the inequality
\be  (\gamma_{\rm s})^2 \leq {54 \over (3N^2-1) }  \times  
\left({g^2 N \over 4 \pi^2}\right)^2 \, .\ee
For the anomalous dimensions of the operators contributing to the 
${\cK}(x)\cdot {\cK}(y)$ OPE we finally find 
\be
\gamma_{\rm min} \leq {3(6 N^2+1) \over 3N^2-2} \times  
{g^2 N \over 4 \pi^2} \leq \gamma_{\rm Max} \, . 
\ee
Again in the limit $N \rightarrow \infty$, $g^2N$ fixed, $\gamma_{\rm Max}$ 
will be non-vanishing, like in the previous case.
Note that two sets of operators contributing to the OPE of
${\cK}(x)\cdot {\cK}(y)$ and ${\cal{Q}}_{\bf 20^\prime}^{(i_1 j_1)}(x)
\cdot {\cal{Q}}_{\bf 20^\prime}^{(i_2 j_2)}(y)$ have a non-trivial 
intersection, since the function $\langle {\cal C} {\cal C}^{\dagger} 
{\cal K}_{\bf 1} {\cal K}_{\bf 1} \rangle$ is non-vanishing, 
but obviously do not to coincide.

We conclude this section by observing that without performing 
further calculations, either at higher orders or involving 
operators with fermionic content, it is impossible to disentangle 
all the $\Delta_0=4$ operators belonging to the singlet and 
the icosaplet representations. One might be tempted to try an ansatz 
that satisfy all the constraints with a number of operators 
smaller than the maximum in principle allowed. Indeed there are 
many such possible choices. Rather surprisingly, it turns out that a 
very severe constraint on any such conjecture comes from the 
requirement that the right hand side of eq.~(\ref{Pnloop}) has 
to be polynomial in $N$ for any $n \leq 3$.

\section{Vanishing of instanton contributions}

We would like to prove the validity of the non-perturbative result
\be
\langle \K1 \K1 \K1 \K1 \rangle_{\rm np} = 0
\label{K4}\ee
for any instanton number and any gauge group.

Let us start considering the case of one-instanton ($\kappa =1$) and
$SU(2)$ gauge group. For this type of calculations it is more
convenient to work in the Wess-Zumino gauge, where the operator ${\cK}$
takes the simple form
\be
\K1 = {1 \over 4}\sum_{i=1}^{6}\sum_{a=1}^{3}\varphi^i_a \varphi^a_i\, .
\label{kappa1}\ee
The effective one-instanton contribution to the fundamental scalars,
$\varphi$, is given by
\be
\varphi^i_a = f^2(x) t^i_{AB} \zeta^A \sigma_a \zeta^B\, ,
\label{phinst}\ee
where
\be
f(x) = {\rho \over (x-x_0)^2 + \rho^2}
\ee
is the instanton profile function (incidentally, in the AdS/CFT
correspondence~\cite{jm,gkp,w} $f(x)$ plays the r\^ole of
boundary-to-bulk propagator~\cite{bgkr}) and
\be
\zeta^A_\alpha = \eta^A_\alpha +\hat{x}_{\alpha\dot\alpha}
\xibar^{A\dot\alpha}
\label{zeta}\ee
is a Weyl spinor in the representation ${\bf 4}$ of $SU(4)$ with
$\hat{x}_{\alpha\dot\alpha} = x_\mu \sigma^\mu_{\alpha\dot\alpha}$.
Explicitly, inserting eqs.~(\ref{phinst}) and~(\ref{zeta})
in~(\ref{kappa1}), we get
\be
\K1\Big{|}_{\rm inst} = f^4(x) \delta_{ij} t^i_{AB} 
(\zeta^A \sigma_a \zeta^B)
t^j_{CD} (\zeta^C \sigma_a \zeta^D)\, .
\label{kapinst}
\ee
Form the completeness relation of $\sigma$ matrices and the $SU(4)$
relation
\be
\delta_{ij} t^i_{AB} t^j_{CD} = 2\epsilon_{ABCD} \, ,
\label{SU4eps}
\ee
one immediately gets the identity
\be
\K1\Big{|}_{\rm inst} = 2 f^4(x)\epsilon_{ABCD} (\zeta^A \zeta^B)
(\zeta^C \zeta^D)=0\, .
\label{kinst}
\ee
The last equality follows from the antisymmetry of the
$\epsilon$-symbol and the symmetry of $(\zeta^A\zeta^B)$.
The vanishing of the one-instanton contribution to the operator $\K1$,
trivially implies for the $SU(2)$ case
\be
\langle \K1 \K1 \K1 \K1 \rangle^{SU(2)}_{\kappa=1} = 0 \, ,
\ee
since there is no way to absorb the 16 exact fermionic zero-modes that
exist in the one-instanton background~\footnote{We recall that the 16
(supersymmetric and superconformal) fermionic zero-modes, generated by
the 16 possible independent choices of the spinors $\eta^A$ and $\bar\xi^A$
in eq.~(\ref{zeta}), are lifted neither by the Yukawa couplings nor by the
quartic potential term of the scalars~\cite{dkm}.}.

The generalization to higher instanton numbers and other gauge groups is
as always straightforward. An instanton correlator may be non-vanishing only if
all the ``unlifted'' 16 supersymmetric and superconformal fermionic
zero-modes are absorbed by suitable operator insertions. But we have just 
shown that $\K1$ can be of no help in this job. Thus the correlator
$\langle \K1 \K1 \K1 \K1 \rangle$ vanishes for any $\kappa$ and any
gauge group. The same type of argument leads to the conclusion that the
three-point function $\langle \Q20 \Q20 \K1 \rangle_{\kappa}$ and
four-point correlators, such as $\langle \Q20 \Q20 \K1 \K1 \rangle_{\kappa}$
or $\langle \Q20 \Q20 \Q20 \K1 \rangle_{\kappa}$,
are zero for any $\kappa$ and any gauge group. Similar results extend 
to other operators in the Konishi supermultiplet. For instance
\be
\langle \calK (x_1) \ldots \calK (x_n) \rangle_{\kappa} = 0
\ee
for any $n\le 4$, any $\kappa$ and any choice of $\calK$ in the
multiplet.

However, five- and higher-point functions with insertions of operators
in the Konishi multiplet can receive non-vanishing contributions from
instantons. The simplest example of a correlator of this type is  
\be \langle \Q20(x_1) \Q20(x_2) \Q20(x_3)
\Q20(x_4) \cK(x_5) \rangle \, ,
\label{5pointdef}
\ee  
Similarly the simplest non-vanishing correlator with only $\cK$ 
insertions is the eight-point function.

Based on the OPE analysis of the one-instanton contribution to the
four-point correlator $\langle \Q20 \Q20 \Q20 \Q20 \rangle$, we proved in
ref.~\cite{bkrs1} the result $\gamma_{{\cK}}^{(\kappa=1)}=0$. The new
calculations presented here seem to point to a much larger ``inertia''
of the Konishi operators to non-perturbative corrections, namely to the
vanishing of the whole non-perturbative correction to their anomalous
dimension ($\gamma_{{\cK}}^{\rm (np)}\equiv 0$) and to the vanishing
of similar corrections to trilinear couplings, like 
$\langle \K1 \K1 \K1 \rangle$ or $\langle \Q20 \Q20 \K1 \rangle$.

\section{Conclusions}

In this concluding section we would like to briefly summarize the present
understanding of the spectrum and properties of composite operators in
${\cal N}=4$ SYM in view of the AdS/CFT correspondence.

First of all, there are short multiplets with maximal spin 2, which are
dual to the supergravity multiplet and its Kaluza-Klein (KK) excitations.
Their scaling dimensions, which are integer for bosons and half-integer
for fermions, do not receive quantum corrections and the same seems to be
true for their trilinear couplings~\cite{lmrs,dfs,bkrs1,bkrs2}. AdS
computations at strong coupling~\cite{dfmmr} have shown that 
extremal~\cite{liutse,af3} and next-to-extremal~\cite{epv}
correlators of these operators do not receive quantum
corrections either. The AdS results have been tested
both in perturbation theory and non-perturbatively with ${\cal{N}}=4$
field theoretical computations~\cite{bk,ehssw,epv}.

Other short multiplets with spin larger than two, that cannot possibly
be dual to the supergravity fields or their KK excitations, are known to
have similar non-renormalization properties~\cite{skiba}. 
Multitrace operators of this kind are interpreted as dual to 
BPS-like bound states at threshold.

Next there are multitrace operators in long multiplets with anomalous
dimensions that vanish in the large $N$ limit and are thus visible in the
supergravity approximation~\cite{dfmmr4,af4}. Their interpretation is more 
troublesome. They can be viewed as non-BPS bound states, since their 
anomalous dimensions are of the correct order of magnitude, 
\ie~$\gamma\propto 1/N^2$, to be holographically dual to gravitational 
binding energies~\footnote{A gravitational-like mass defect 
$\delta{M} = {G_{N}M^{2}/ L^{2}}\approx g_{s}^{2}/L = {g_{YM}^{4} 
N^{2}/ N^{2} L}$ precisely corresponds to an anomalous dimension 
$\gamma \propto {1/N^{2}}$ at fixed `t Hooft coupling.},
but it is not clear what is the meaning that should be attached to a
``classical'' bound state~\footnote{We would like to thank A.C. Petkou
for an interesting discussion on this point.}.

We then have long Konishi-like multiplets that are expected to
acquire large anomalous dimensions ($\Delta \approx (g^2 N)^{1/4}$) and
decouple from the operator algebra at strong 't~Hooft coupling in the
large $N$ limit. They are dual to string excitations, whose mass is of the
order of $\Delta$ in AdS units. In this paper we have confirmed
previous results on the perturbative anomalous dimension of the ${\cal
N}=4$ Konishi multiplet and gone one step further on the
non-perturbative side by showing the vanishing of instanton
contributions to correlators with up to four operator insertions.

Finally there are unprotected operators that rather surprisingly do
not receive corrections to their tree-level dimensions. This was known
for a double trace operator of dimension $\Delta_0=4$ belonging to the 
representation ${\bf 20^\prime}$ of $SU(4)$ at O($g^2$) and including 
one-instanton corrections~\cite{afp}. 
We have confirmed this at O($g^4$) (see also~\cite{aeps}) and for 
any winding number in the present paper. Partial non-renormalization 
of near-extremal correlators of CPO's~\cite{epss}, and in particular a 
functional relation between two {\it a priori} independent 
four-point functions of lowest dimension CPO's~\cite{bkrs2,ess}, seems 
to be at the heart of the vanishing of the anomalous dimensions of some 
unprotected operators. It would be interesting to better understand the
constraints imposed by ${\cal N}=4$ superconformal invariance along the
lines of what has been done for four-point functions in~\cite{dolosb}.

We would like to conclude with a few comments on the $U_B(1)$ 
bonus symmetry of ref.~\cite{intri} and on the $SL(2,Z)$ invariance 
expected to be realized in ${\cal N}=4$ SYM. 

On the first issue our results confirm the conjecture that correlators 
of single-trace operators in short multiplets with up to four-point 
as well as three-point functions with at most one operator belonging 
to a long multiplet obey the $U_B(1)$ selection rule.

As for the question of how $SL(2,Z)$ invariance is realized, we can make 
the following observations. Since physics is $\theta$ dependent in 
${\cal N}=4$ SYM, we expect 
observables, such as anomalous dimensions and trilinear couplings, to
generally depend on the vacuum angle. It is thus conceivable that
some of the observables may be non-holomorphic, but still $\theta$ 
dependent, modular functions of the
complexified coupling, $\tau = \theta/2\pi + 4\pi i/g^{2}$. 
The class of operators whose dimensions and couplings may depend
on $\tau$ in a modular invariant way is, however, restricted to the operators 
with $\gamma\propto 1/N^2$ in the large $N$
limit, like, for instance, the unprotected double-trace operators 
of dimension 4 in the singlet of $SU(4)$. As remarked above, when 
discussing the relation between gravitational binding energy of non 
BPS states and anomalous dimensions, this kind of non-BPS multi-trace 
operators (dual to non BPS multi-particle states) are rather ``elusive'' 
from the AdS perspective. Konishi-like operators, on the contrary, do 
not seem to receive any non-perturbative corrections to their anomalous
dimensions and trilinear couplings. This means that none of the observable
quantities associated to them can possibly show simple modular properties.  
A better understanding of string theory on AdS space and in general on 
backgrounds with non-vanishing RR charge might shed some light 
on the pattern of dimensions of operators dual to string excitations 
and non-BPS bound states of KK excitations. 

In conclusion ${\cal N}=4$ SYM seems to be a very interesting theory.
Though superconformal symmetry strongly constrains the dynamics, it
allows for interesting features to emerge both at weak coupling (in
perturbation theory and non-perturbatively) and at strong coupling,
where the supergravity description is in good qualitative agreement
with field theory expectations~\cite{mb}.

Some puzzling features call for a deeper understanding of 
the r\^ole of superconformal invariance on the dynamics or for the 
emergence of some new hidden symmetries, such as the
$U_B(1)$ bonus symmetry or symmetries associated to ``central''
extensions of the superconformal algebra.

\section*{Acknowledgements}

We would like to acknowledge stimulating discussions with B. Eden, S. Ferrara, 
D. Freedman, M.B. Green, P. Howe, H. Osborn, A.C. Petkou, A. Sagnotti, 
K. Skenderis, E. Sokatchev and G.Veneziano. This work was partly supported 
by the EEC contract HPRN-CT-2000-00122. M.B. also ackowledges partial 
financial support by PPARC and the INTAS project 991590.

\end{document}